\def\year{2022}\relax
\documentclass[letterpaper]{article} 
\usepackage{aaai22}  
\usepackage{times}  
\usepackage{helvet} 
\usepackage{courier}  
\usepackage[hyphens]{url}  
\usepackage{graphicx} 
\usepackage{natbib}  
\usepackage{caption} 
\usepackage{booktabs}

\title{The Contribution of Lyrics and Acoustics to Collaborative Understanding of Mood}
\author {
    Shahrzad Naseri\textsuperscript{\rm 1}\thanks{The research was conducted while these authors were at Spotify.},
    Sravana Reddy\textsuperscript{\rm 2}\footnotemark[1],
    Joana Correia\textsuperscript{\rm 3},
    Jussi Karlgren\textsuperscript{\rm 3}, 
    Rosie Jones\textsuperscript{\rm 3} \\
}
\affiliations{
     \textsuperscript{\rm 1} University of Massachusetts Amherst \\
     \textsuperscript{\rm 2} ASAPP \\
     \textsuperscript{\rm 3} Spotify Research \\
    shnaseri@cs.umass.edu, sravana.reddy@gmail.com, \{joanac, jkarlgren, rjones\}@spotify.com
}

\usepackage{amsmath}
\usepackage{amssymb}
\usepackage{graphicx}
\graphicspath{ {./figures/} }
\usepackage{booktabs}
\usepackage[htt]{hyphenat}
\usepackage[dvipsnames]{xcolor}
\usepackage{multirow}
\usepackage{appendix}

\begin{document}

\maketitle

\begin{abstract}
In this work, we study the association between song lyrics and mood through a data-driven analysis. Our data set consists of nearly one million songs, with song-mood associations derived from user playlists on the Spotify streaming platform. We take advantage of state-of-the-art natural language processing models based on transformers to learn the association between the lyrics and moods. We find that a pretrained transformer-based language model in a zero-shot setting -- i.e., out of the box with no further training on our data -- is powerful for capturing song-mood associations. Moreover, we illustrate that training on song-mood associations results in a highly accurate model that predicts these associations for unseen songs. Furthermore, by comparing the prediction of a model using lyrics with one using acoustic features, we observe that the relative importance of lyrics for mood prediction in comparison with acoustics depends on the specific mood. Finally, we verify if the models are capturing the same information about lyrics and acoustics as humans through an annotation task where we obtain human judgments of mood-song relevance based on lyrics and acoustics.  
\end{abstract}

\section{Introduction}\label{sec:intro}

Lyrics are important for the musical experience, providing us with the rich stories and messages that artists want to convey through their music. 
However, the perceived mood of a song may not stem from its lyrics alone. 
Take, for example, the song {\em Cardigan} by Taylor Swift:

\begin{quote}
\textit{'Cause I knew you}\\
\textit{Steppin' on the last train}\\
\textit{Marked me like a bloodstain, I}\\
\textit{I knew you}\\
\textit{Tried to change the ending}\\
\textit{Peter losing Wendy, I}\\
\textit{I knew you}\\
\end{quote}

The lyrics are about the end of a relationship, and imply moods of sadness, longing, and heartbreak. However, the acoustic mood of the song with its familiar chord progressions and somewhat high tempo is calm and upbeat. This example is not an exception; in fact, a recent analysis of the lyrics and acoustics of popular music \cite{interiano2018musical} identifies a trend where songs have been getting sadder lyrically in the last three decades, but also more `danceable' and `relaxed.'

In this paper, we investigate the association between song lyrics and \textit{moods} -- terms that describe affectual qualities of a song -- and conduct a data driven analysis using state of the art natural language processing models to compare how lyrics  contribute to the understanding of mood as defined collaboratively by the playlisting behavior of users of the Spotify music streaming platform.

Previous studies in the psychology of music have shown that acoustics and lyrics play different roles in listener perceptions, and that these roles depend on the specific mood. One study \cite{doi:10.1177/0305735606067168} showed that lyrics detract from emotion in happy and calm music in their participants, but enhance emotion in sad and angry music. Results from an fMRI study \cite{10.3389/fpsyg.2011.00308} lend support to the hypothesis that  lyrics are more important for the perception of sad emotions than acoustics, but that acoustics are of primary importance for the perception of happy emotions. 

One motivation for our project is to tackle the above questions of music perception outside the laboratory, from the perspective of a large-scale data set of music originally tagged with moods by listeners on the Spotify music streaming platform.  
The second motivation is to explore how machine learning models might be designed to automatically associate moods with songs in order to  enable listeners to search and discover music. While previous work \cite{Zaanen2010AutomaticMC,laurier2008multimodal,hu2010lyrics} 
has explored this modeling problem, our study uses a data set that is orders of magnitude larger. In addition, while previous work largely relies on bag-of-words models, we use state of the art natural language processing models based on the transformer architecture \cite{vaswani2017attention} to better capture the semantic nuances of lyrics in order to learn the associations between lyrics and moods.


We break down our overarching research question, \textit{``How much do lyrics and acoustics of a song each contribute to understanding of the song's mood?"}, into the following sub-questions:
\begin{description}
\item[RQ1] What can lyrics tell us about moods with {\em no} training on lyric-mood associations?
\item[RQ2] Can training a lyrics-based model on listener-generated mood tags  produce accurate mood associations?
\item[RQ3] How much do lyrics contribute to moods compared to acoustics?
\item[RQ4] Do models capture the same information about lyrics and acoustics as humans?
\end{description}

In order to answer these questions, we take advantage of a transformer model to represent lyrics as well as bag-of-words representations. The results show that using a zero-shot model (i.e., a model pretrained on web data with no training on our data set)
is powerful in capturing the mood associations of songs. 
Training on listener-generated mood-song associations results in a model that captures these associations more accurately. 

By comparing the prediction of models based on lyrics and acoustics, we find that the contribution of lyrics varies depending on the specific mood. We observe a similar result on conducting a manual annotation task  where we elicit judgments on mood-song relevance based on lyrics and acoustics in isolation. 



\section{Mood in Music and Text: A Brief Survey}

\subsection{Music Psychology}
The study of how humans connect moods and music has been explored through experiments in the fields of psychology and neuroscience. 
 One study 
 \cite{doi:10.1177/0305735606067168} finds that acoustics are more dominant than lyrics in eliciting emotions in study participants, and that lyrics play a bigger role in perceived sad or angry music compared to music with positive moods. This is similar to the finding of an fMRI experiment \cite{10.3389/fpsyg.2011.00308} showing that lyrics define moods for sad music whereas acoustics are of primary importance for happy moods.
 The relative unimportance of lyrics on mood compared to acoustics is corroborated by other studies \cite{doi:10.2466/pms.1997.85.1.31}, but contradicted by others: for example, one paper \cite{doi:10.2190/35T0-U4DT-N09Q-LQHW}  finds that sad lyrics with upbeat acoustics elicits negative emotion, and yet another contradicting paper
\cite{doi:10.1177/0305735613483667} demonstrates that sad lyrics seem to enhance pleasant feelings induced by happy-sounding music compared the acoustics alone. Finally, it has been shown \cite{doi:10.1111/1467-9280.00091} that humans appear to process song melodies and lyrics independently, raising questions about how conflicting moods in the acoustics and lyrics are processed. 

One should keep in mind that some of the contradicting results from these laboratory studies may be due to variations in experimental design, the selection of participants and the music, small sample sizes, and the time period or cultural landscape in which the studies were conducted. We believe that a computational lens on these questions using web-scale data, while not perfect, is a complementary angle to such works. 

\subsection{Prediction of Mood with Lyrics and Acoustics}
The annual Annual Music Information Retrieval Evaluation eXchange (MIREX) introduced a music mood classification task in 2007 \cite{downie20082007}. This task explicitly disallowed consideration of lyrics in classification or evaluation. Submitted models were found to have overall better classification performance using acoustics for mood clusters like \{`wistful', `brooding'\} and \{`volatile', `fiery'\} compared to clusters like \{`rousing', `confident'\} and \{`fun', `cheerful'\}.
\citeauthor{laurier2008multimodal} (\citeyear{laurier2008multimodal}) collect a 1000-song and 4-mood data set of song-mood associations by eliciting judgments from annotators, and build classifiers using bag-of-words and latent semantic analysis features for lyrics, and acoustic features like timbre, tempo, and pitch. They find that acoustics and lyrics combined improve prediction of `happy' and `sad' moods, but that the performance is relatively saturated with acoustics alone for `anger.' A follow-up study \cite{hu2010lyrics} with a slightly larger data set of a few thousand songs and 18 moods finds that lyrics significantly outperform acoustics for prediction. A smaller scale and more detailed approach \cite{schmidt2011modeling} models the temporal dynamics of emotional approaches to songs.

Since annotations of mood for large collections of music are hard to obtain, \citeauthor{mcvicar2011mining} (\citeyear{mcvicar2011mining}) use unsupervised methods to correlate acoustic attributes and lyrics and find through canonical correlation analysis that the top correlation components correspond to mood, suggesting that lyrics and acoustic attributes in a song tend to have consistent moods overall.

Recently, there have been a body of works that applied deep neural network models to capture the association of mood/emotion and song by taking advantage of audio features~\cite{saari2013role, panda2019emotion, korzeniowski2020mood, panda2020audio, medina2020emotional}, lyrics features~\cite{fell2019love, hrustanovic2021recognition} as well as both lyrics and audio~\cite{delbouys2018music, parisi2019exploiting, wang2021method, bhattacharya2018multimodal} features.  Delbouys et al. classify mood of a song to either `arousal' or `valence' by utilizing a 100-dimensional word2vec embedding vector that is trained on 1.6 million lyrics in 
several different neural architectures such as GRU, LSTM, Convolutional Networks for their lyrics-based model. Further, they utilize  audio mel-spectrogram as input to a  convolutional neural network model. Parisi et al. show a comparison between text-based and audio-based deep learning classification models to classify the mood of a song to 5 discrete crowd-based adjectives: \{`sad', `joy', `fear', `anger', `disgust'\}. 

\subsection{Sentiment Analysis of Language}

Some papers \cite[etc]{5363083,Zaanen2010AutomaticMC,fell2014lyrics} have used lyrics with bag-of-words based models to predict mood without comparison to acoustics. An analysis of proxies for linguistic creativity \cite{hu2011exploring} shows that sad or negative lyrics score higher in creativity than positive songs. 

Outside of music, the detection of mood and affect in human language use more generally has been the object of systematic computational study since the first AAAI Symposium on the topic \citep{qu2004aaai} and traces its modern beginnings to the study of human emotional expression by Charles Darwin (\citeyear{darwin}) and others \cite[e.g.]{james1884}. 

Sentiment analysis models have historically been based on bag of words classifiers and lexical look-ups, with some syntactical finesse to process e.g. negation or amplification \citep{pang2008opinion}. This type of approach is deterministic and interpretable but at a cost for coverage. Deep learning approaches that can model contextual relationships have been showing improvements over traditional methods \citep{zhang2018deep}. In particular, the use of pretrained transformer models such as BERT \cite{devlin-etal-2019-bert} have seen success in recent years \cite{li-etal-2019-exploiting, yin-etal-2020-sentibert}. However, these models have not been applied to music or mood categories to the best of our knowledge.


\section{Data}
Our study analyzes the association of a song's lyrics with the set of terms describing its mood. Our palette of moods consists of 287 terms in English. 
 This  set of moods includes terms like ``chill'', ``sad'', ``happy'', ``love', and ``exciting''. The moods are not limited to a specific part-of-speech, covering not only adjectives (``sad'', ``somber'', etc), but also nouns (``motivation'', ``love'', etc.) and verbs (``reminisce'', ``fantasize'', etc.). 


The association between a song and the mood is calculated using collaborative data (by ``wisdom of the crowd''). More specifically, Spotify music streaming platform provides playlists of songs as well as enabling users to create their own playlists. The playlists have a name and an optional description. To calculate the association between song and mood, the Spotify music streaming platform starts from their collection of
($\approx$4 billion) playlists 
and filters down to those
playlists that have words corresponding to the mood lexicon in their titles and/or descriptions.
The co-occurrence between each mood and song is then computed. Finally, the association~\cite{wang2021systems} between a mood and a song is calculated according to the Pointwise Mutual Information (PMI):

\begin{align}
    \text{PMI}(s,m) = \log\frac{p(s,m)}{p(s)p(m)}
\end{align}

where $s$ and $m$ are the songs and moods, respectively. $p(s,m)$ is the probability of co-occurrence of the song and mood. The PMI score shows how much more likely is the song $s$ to co-occur with the mood $m$. 

The final association score is a slight variation of the PMI score called Bayesian Normalized Pointwise Mutual Information (BNPMI) where instead of using empirical probability of the probability of a song given a mood,
a conjugate beta prior~\cite{schlaifer1961applied} with parameters estimated by method of moments~\cite{hall2005generalized} is used. This change compensates for the rarer song and moods that \textit{coincidentally} are co-occurring or not co-occurring. 

BNPMI scores are between [-1,+1], where -1 represents negative association (the mood and song {\em never} co-occur), 0 of independence, and +1 of perfect co-occurrence. 
For more details about the BNPMI score, see the Appendix.


Associating songs and moods through terms in user playlists is not as precise as explicitly eliciting mood tags, but avoids biases in elicitation, and allows us to scale massively to a large number of songs, with the association collected from millions of diverse users.
Our data set contains $\approx$ 955K songs. The lyrics of these songs are obtained from a commercial service\footnote{\url{https://www.musixmatch.com}} which precludes the sharing and distribution of these specific data as a separate collection. However, to give a sense of our dataset we provide 18 pairs of (song, mood) for 3 different songs in the Appendix. We break down the songs into train and test sets by reserving 75\% and 25\% of samples for train and test sets, respectively.  

Moreover, in the process of creating the dataset we run into the duplicate (song, lyrics) tuples because of different spacing and new line insertions in the lyrics text. Since each song has a unique identifier, we only select one of the instances of duplicate tuples with the same song identifier. 

\begin{figure}[h]
\centering
\includegraphics[scale=0.3]{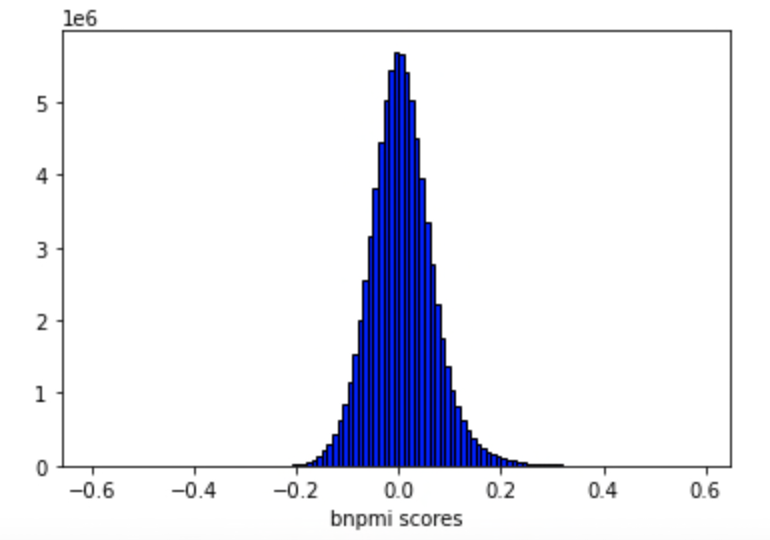}
\caption{BNPMI score distribution}
\label{fig:bnpmi_dist}
\end{figure}

For training and evaluation of our models, we label the association between a song and a mood by binning the BNPMI scores.  
More specifically, if the BNPMI score is greater than the threshold $\tau$ it represents a positive association and if it's lower than the $-\tau$ it represents a negative association. The values that fall between $-\tau$ and $\tau$ correspond to a neutral association. Figure~\ref{fig:bnpmi_dist} shows the histogram of BNPMI scores which is a normal distribution with mean equal to -0.0034 and standard deviation equal to 0.0861. In our experiments, we select the threshold $\tau=0.1$, which the neutral association will fall within approximately one standard deviation of the mean.
 
Limiting the association only to positive and negative association results in $\approx$ 2 million (song, mood) pairs for training and $\approx$ 774K pairs for testing. The exact number of instances is shown in table~\ref{tab:train_test_validation}. Moreover, we do not perform any negative sampling and we utilized all of the negative bnpmi score association in training our model.

\begin{figure*}[h]
\centering
\includegraphics[scale=0.58]{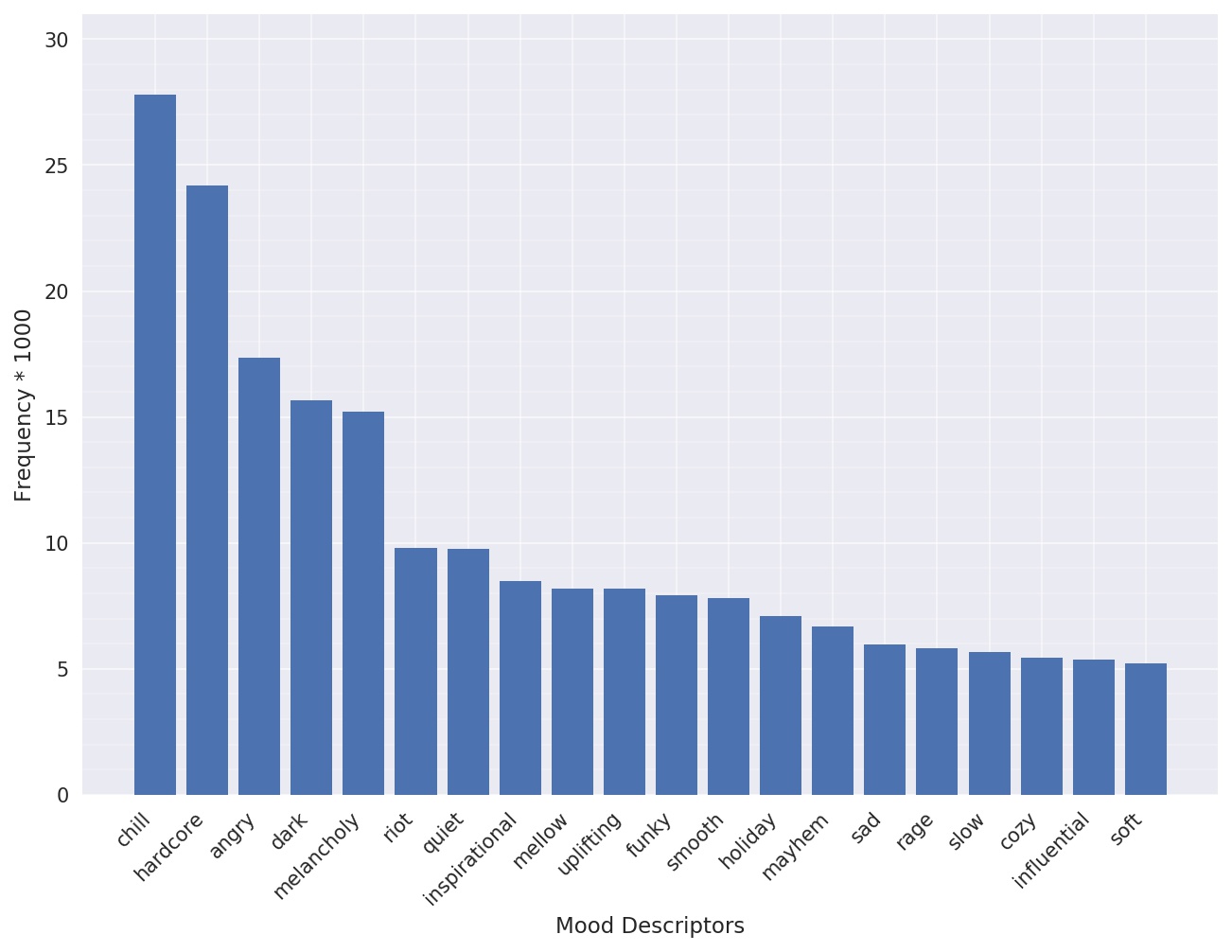}
\caption{Frequency of top 20 mood descriptors with positive association to their corresponding song in the test split.}
\label{fig:freq_positive_desc_test_data}
\end{figure*}

Lastly, Figure~\ref{fig:freq_positive_desc_test_data} shows the frequency of top 20 mood descriptors with positive association to their corresponding song. ``Chill" mood descriptors is the most frequent mood positively associated to songs. 

\begin{table}[t]
    \centering
    \scalebox{1}{
    \begin{tabular}{c ccc}
        \toprule
                           & \textbf{Total} & \textbf{Train} & \textbf{Test} \\
        \cmidrule{2-4}
        \textbf{song} &  955,109     & 716,331        & 238,778       \\
        \textbf{(song, mood)} & 3,083,727 & 2,309,083 & 774,644 \\
         \bottomrule
    \end{tabular}
    }
    \caption{Number of song and (song, mood) pairs.}
    \label{tab:train_test_validation}
\end{table}

\section{Methodology and Experiments}
In this section, we empirically investigate the answer to the overarching research question: ``How much the lyrics and the acoustics of a song each contribute to understanding of the song's mood?''. 


\subsection{RQ1. What Can Lyrics Tell us about Moods with {\em No} Training on Lyric-Mood Associations?}
\label{sec:zeroshot}

To answer this question, we formulate the problem as a zero-shot text classification task in which the model predicts the class of a text without having seen a single labeled example, instead using linguistic knowledge from a model pretrained on a different domain.
Natural Language Inference (NLI)~\cite{bowman2015large}, the task of determining whether or not a premise sentence semantically entails a hypothesis sentence, is used as a proxy to tackle the problem of zero-shot text classification of moods. 
This approach 
casts the target label into a sentence and judges whether or not the input text entails the target sentence.

We take advantage of the BART~\cite{lewis2019bart} model that is trained for the NLI task on the Multi-genre Natural Language Inference (2017) corpus~\cite{N18-1101} as our model for the zero-shot text classification task, according to the approach proposed by ~\citeauthor{yin2019benchmarking} (\citeyear{yin2019benchmarking}). The BART model fine-tuned on the MNLI corpus is a transformer-based model that has a classification head with 3 classes: Contradiction, Neutral, and Entailment. 

Under the NLI formulation, lyrics are inputs and moods are targets. The BART model is used without any further training. We cast the mood terms into sentences using templates according to their parts of speech, with manual re-annotations for corner cases. Table~\ref{tab:desc_template} shows the examples of some moods and the corresponding sentences.

\begin{table*}
    \centering
    \begin{tabular}{l cl} 
        \toprule
        \textbf{Mood} &  Part of Speech &  Sentence \\
        \midrule
        Chill & adjective & This is a \textbf{\underline{chill}} song. \\
        Heartbroken & adjective & This song is about being \textbf{\underline{heartbroken}}. \\
        Love & noun & This song is about \textbf{\underline{love}}. \\
        Fantasize & verb & This song makes you \textbf{\underline{fantasize}}. \\
        \bottomrule
    \end{tabular}
    \caption{Mood terms rewritten as sentences}
    \label{tab:desc_template}
\end{table*}

Following \citeauthor{yin2019benchmarking}, the output for the Neutral class is ignored. To be more specific, the association between the lyrics and each mood is formulated as a binary classification task and the final probability of association is calculated according to the Contradiction and Entailment class probabilities.

We evaluate our zero-shot NLI approach against the binned association labels derived according to the BNPMI score (described in the Data section) in Table~\ref{tab:mode_eval_against_bnpmi}. We observe that the zero-shot NLI model has high precision, however it under-predicts the positive association between a song and mood, resulting in low recall. 

We further illustrate the performance of the zero-shot classifier with the example song shown earlier: {\em Cardigan} by Taylor Swift. The listener-generated mood associations as characterized by high BNPMI scores include ``heartbroken'', ``calm'', ``sad'', ``bittersweet'', ``vulnerable'', and ``obsessed''. The classifier predicts all the above moods except ``calm'' to be associated with the song. Note that while our evaluation penalizes the classifier for missing ``calm'', it is in fact doing the right thing since that mood is not consistent with the lyrics, and is presumably derived from acoustic perceptions. 
RQ3 will address this aspect of the evaluations.

\citeauthor{ma-etal-2021-issues} (\citeyear{ma-etal-2021-issues}) recently pointed out a few issues with the NLI approach for tackling the zero-shot text classification problem; in particular, they suggest that NLI is not a good proxy for text classification. They propose instead using models like BERT that are fine-tuned for the task of \textit{next sentence prediction} (NSP), with the reasoning that NSP is a closer proxy of text classification than NLI.

Next sentence prediction is a binary classification task that given two sentences, predicts whether the second sentence follows the first input `sentence'. As before, we use lyrics as the first sentence, and the moods cast as sentences as the second. We can see from the top row of Table~\ref{tab:mode_eval_against_bnpmi} that the F1 score for the NSP-Zeroshot model is indeed higher than MNLI-Zeroshot. However, a closer examination of the confusion matrix shows that the model is biased toward predicting the positive class, resulting in lower precision. 

\paragraph{Takeaways:} Pretrained transformer models are much better than chance at predicting the association between song lyrics and moods, despite being trained on completely different tasks and domains. Next sentence prediction is more effective than natural language inference as a proxy for song-mood association prediction, but the latter shows higher precision. 
We hypothesize that fine-tuning the models on training data of song-mood associations will result in higher performance.

\begin{table*}[t]
    \centering
    {
    \begin{tabular}{lll ccc }
        \toprule
        \textbf{Modality} & \textbf{Approach} & \textbf{Lyrics Model} & Precision & Recall & F1 \\
        \midrule
        
        \multirow{5}{*}{\textbf{Lyrics}} & \multirow{2}{*}{Zero-shot Learning} &  NLI & 83.14 & 43.95 & 57.50 \\
                                                    & & NSP  & 75.51 & 98.22 & 85.38 \\
                                        \cmidrule{2-6}   
                                        & \multirow{4}{*}{Fine-tuned }& BoW  & 93.85 & 90.35 & 91.74 \\
                                                    & & NLI (w/o. Neutral)& 96.48 & 97.63 & 97.05 \\
                                                    & & NLI (w. Neutral) & 95.90 & 96.79 & 96.34\\
                                                    & & NSP  & 96.17 & 97.46 & 96.81 \\

        \midrule
        \textbf{Acoustics} & Fine-tuned  & - & 95.64 & 81.54 & 87.48 \\
        \midrule
        \multirow{2}{*}{\textbf{Hybrid}} & \multirow{2}{*}{Fine-tuned} & BoW  & 95.53 & 94.54 & 94.86 \\
                                                                              & & NLI & 96.57 & 97.93 & 97.24 \\ 
        \bottomrule
    \end{tabular}
    }
    \caption{Lyrics-based, acoustics-based and hybrid models evaluated against the ground truth song-mood association.}
    \label{tab:mode_eval_against_bnpmi}
\end{table*}


\subsection{RQ2.  Can Training a Lyrics-based Model on Listener-generated Mood Tags Produce Accurate Mood Associations?}
To address this question, we take two modeling approaches. Following previous literature in mood predictions from lyrics, we represent the lyrics by a bag-of-words model. We also use a transformer-based model. 
We train both the bag-of-words and transformer-based models on the (lyrics, mood) pairs with the BNPMI-based binned association as the target labels. 

\subsubsection{Bag of Words (BoW)}\label{sec:bow} Lyrics are represented by the $tf.idf$ scores of their unigrams, estimated on the  training set. For each mood, we train a binary logistic regression classifier, classifying whether the mood and the song are positively associated with each other. 

\subsubsection{Transformers} We fine-tune the BART model pre-trained on the MNLI corpus (`fine-tuned NLI') as well as a BERT model trained on the next sentence prediction task (fine-tuned NSP).  Similarly to the zero-shot learning setup in the previous section, the input is a pair of texts, with  the model is classifying whether the first text entails the second, while in the next sentence prediction the model predict if the second text would follow the first in a corpus. The models are trained on the BNPMI-based associations. Table~\ref{tab:bnpmi_based_association} shows the label mappings used for each model.

We fine-tune the NLI model in two ways: excluding and including the neutral moods. Since the number of the neutral mood descriptors is high for each recording, we consider only one neutral mood descriptor for each recording to make the fine-tuning feasible (in the defined train set the total number of recordings with neutral descriptors is approximately 20M pairs).

The results for the BoW, fine-tuned NLI and fine-tuned NSP model are shown in the second row of table~\ref{tab:mode_eval_against_bnpmi}. We observe that fine-tuning shows large improvements in both precision and compared to the zero-shot approach. While the BoW model performs well with a F1 score of 91.74, the transformer models perform better, especially for recall. We also see that the difference between the fine-tuned NLI models (both w/o and w. neutral) and the fine-tuned NSP model is negligible, illustrating that training on listener-generated song-mood associations provides both models with powerful signals that were missing in the zero-shot approach. Moreover, we observe fine-tuning with neutral descriptors decrease the performance across all measures, however this drop is negligible and therefore, throughout the rest of the paper NLI model refers to the NLI (w/o. Neutral) fine-tuning scheme. 

\paragraph{Takeaways:} Overall, fine-tuning on the training data of song-mood associations results in models with high precision and recall that can be valuable for predicting listener-generated moods of new songs, which could then be leveraged for conversational search and recommendation. 

 \begin{table}[t]
    \centering
    {
    \begin{tabular}{l c cc} 
        \toprule
        \textbf{Model} & \textbf{BNPMI Score} & \textbf{Association} & 
        \\
        \midrule
        \multirow{3}{*}{\textbf{NLI}} & [-1, -0.1]  & Contradiction 
        \\
                                      & (-0.1, 0.1) & Neutral  &  \\
                                      & [0.1, 1]    & Entailment
                                      \\
        \midrule
        \multirow{2}{*}{\textbf{NSP}} & [-1, -0.1]  & NotNextSentence 
        \\
                                      & [0.1, 1]    & IsNextSentence  
                                      \\
        \bottomrule
    \end{tabular}
    }
    \caption{Model targets derived from the BNPMI scores.}
    \label{tab:bnpmi_based_association}
\end{table}

\subsection{RQ3. How Much Do Lyrics Contribute to Moods Compared to Acoustics?}

To understand the contribution of the acoustics features to a piece of music's mood, we use a logistic regression classifier with precomputed acoustics features of a song. We also build  hybrid models which take advantage of both lyrics and acoustics. 

\subsubsection{Acoustics} The model represents each song by a set of numerical features corresponding to the acoustics of the song as provided by Spotify API. 
Table~\ref{tab:acoustics_features} shows the list of acoustic features. Similarly to the bag of words model, we train a binary logistic regression classifier for each mood to classify whether the mood and the song's acoustics are associated with each other. 

The third row in Table~\ref{tab:mode_eval_against_bnpmi} shows the results of this model. We observe that the acoustic model does better than the zero-shot approaches with respect to precision, and in the case of the NLI model with respect to recall as well; however, it is clearly outperformed by the fine-tuned lyrics models with respect to recall, where the lyrics are considerably better for coverage. 

In the example of the {\em Cardigan} song, we find that the acoustics model predicts the mood ``calm'', which is associated with the song according to listener-generated playlist data, but is not predicted by the zero-shot lyrics classifier.

\begin{table*}[t]
    \centering
    {
    \begin{tabular}{rl}
        \toprule
        \textbf{Acoustic Features} & acousticness, bounciness, beat strength, danceability, energy, \\
        & flatness, instrumentalness, liveness, loudness, longest silence ratio, \\
        & mechanism, organism, runnability, speechiness, tempo, valence, mean of dynamic range\\
        \bottomrule
    \end{tabular}
    }
    \caption{The list of numerical acoustic features provided by the Spotify API.}
    \label{tab:acoustics_features}
\end{table*}

\subsubsection{Hybrid}
We investigate two hybrid lyrics+acoustics models -- representing a song based on both its lyrics and acoustics features. One uses the bag of words representation of the lyrics (`Hybrid-BoW') and the other uses the NLI transformer-based representation (`Hybrid-NLI'). It is worth mentioning since the difference between the performance of fine-tuned NSP and fine-tuned NLI models is negligible, we only select the NLI model to be used in our hybrid architecture for transformer-based representation of the lyrics.

\textbf{Hybrid-BoW} represents the lyrics and acoustics features by a concatenation of the $tf.idf$ vectors of lyrics and the acoustic features.
It is trained the same way as the bag of words and acoustic models.

\begin{figure*}[h]
\centering
\includegraphics[scale=0.58]{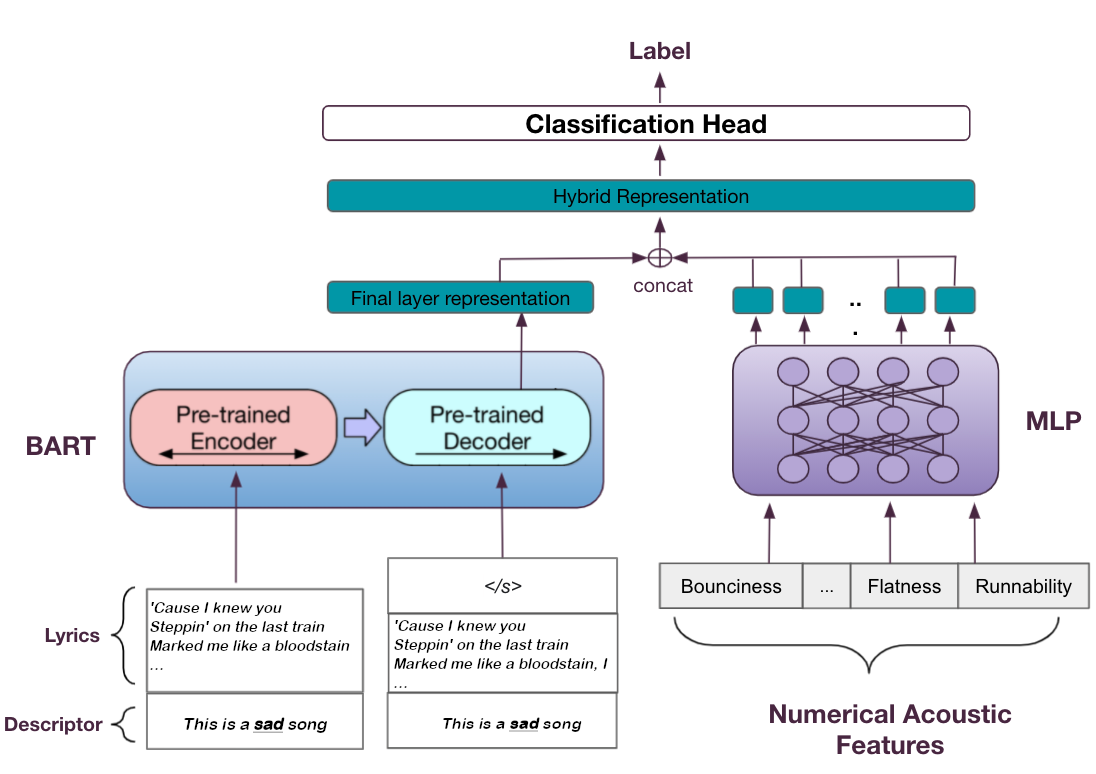}
\caption{Architecture of the Hybrid NLI model.}
\label{fig:hybrid_nli_arch}
\end{figure*}

\textbf{Hybrid-NLI} uses the final layer before the classification layer of the BART transformer fine-tuned on the NLI task to represent lyrics. The acoustic features are fed into a multilayer perceptron (MLP) model to construct the acoustic representation. The lyrics and  acoustic representations are concatenated, which result in the hybrid representation of the song. The hybrid representation is input to a classification head, similar to the classification head in the BART model fine-tuned on the sequence classification task\footnote{ \url{https://huggingface.co/transformers/v2.11.0/_modules/transformers/modeling_bart.html\#BartForSequenceClassification}}. Figure~\ref{fig:hybrid_nli_arch} shows the architecture of the Hybrid-NLI model. 

We compare the results of the hybrid models to the lyrics-only and acoustics-only models in the final rows of Table~\ref{tab:mode_eval_against_bnpmi}. We observe that the Hybrid-BoW model has better recall than the acoustics model and is overall better than the fine-tuned BoW representation of the lyrics. The Hybrid-NLI model behaves similarly to the Hybrid-BoW model, outperforming acoustics-based and lyrics-based models in both recall and precision. The Hybrid-NLI model performance compared to the Hybrid-BoW model highlights the effectiveness of a transformer-based representation of the lyrics. 

To better understand the behaviour of the lyrics-based, acoustics-based, and hybrid models for different moods, we compare the performance of these models for two selected moods, ``chill'' and ``love'', in Table~\ref{tab:case_study}. We observe that the fine-tuned NLI lyrics-based model outperforms the zero-shot and fine-tuned bag of words lyrics-based and acoustics-based models for both cases. Notably, lyrics-based models consistently perform worse for ``chill'' than they do for ``love''; for ``love'', the lyrics alone do better than the acoustics alone. We can see the Hybrid-NLI model does best of all models for ``chill'' and  has competitive performance with the fine-tuned NLI model for ``love''. 

\paragraph{Takeaways:}
These results show that some listener-generated mood tags, e.g. ``chill'', are better predicted from acoustics than from lyrics, in contrast to other mood tags such as ``love''. 
Table~\ref{tab:case_study} suggests that listeners may be paying differential attention to lyrics and acoustics when describing playlists with different moods, and that the difference between the semantics of ``love'' and ``chill'' plays a role here: the latter term is arguably less specific than the former and can thus be assumed to be realised with a larger variety of expression in lyrics, some of which may have to with temperature and meteorological considerations rather than mood and emotion. This polysemy is reduced by the introduction of acoustic features and through fine-tuning the model, which is evident in the scores given in the table. 

Next, in RQ4, we will show that these results are consistent with human judgments in an annotation study. 

\begin{table*}[t]
    \centering
    {
    \begin{tabular}{llll cccc} 
        \toprule
        \textbf{Mood} & \textbf{Feature} & \textbf{Approach} & \textbf{Lyrics Model} & Precision & Recall & F1\\
        \midrule
        \multirow{6}{*}{\textbf{Love}}    & \multirow{3}{*}{\textbf{Lyrics}} & Zero-shot Learning                   & NLI & 77.62 & 86.82 & 81.97 \\
                                                                             \cmidrule{3-7}
                                          &                                  & \multirow{2}{*}{Fine-tuned on BNPMI} & BoW & 87.11 & 83.70 & 85.37 \\
                                          &                                  &                                      & NLI & 92.73 & 90.87 & 91.79 \\
                                          \cmidrule{2-7}
                                          & \textbf{Acoustics}               & Fine-tuned                   &    & 76.67 & 82.19 & 79.33 \\
                                          \cmidrule{2-7}
                                          & \textbf{Hybrid}                  & \multirow{2}{*}{Fine-tuned } & BoW & 89.13 & 87.13 & 88.12 \\
                                          &                                  &                                      & NLI & 93.04 & 90.36 & 91.68
\\
        \midrule
        \multirow{6}{*}{\textbf{Chill}}   & \multirow{3}{*}{\textbf{Lyrics}} & Zero-shot Learning                   & NLI & 11.53 & 7.98  & 9.43  \\
                                                                             \cmidrule{3-7}
                                          &                                  & \multirow{2}{*}{Fine-tuned } & BoW & 27.92 & 78.19 & 41.15 \\
                                          &                                  &                                      & NLI & 80.84 & 67.43 &	73.53 \\
                                          \cmidrule{2-7}
                                          & \textbf{Acoustics}               & Fine-tuned                   & -   & 39.84 & 88.09 & 54.86 \\
                                          \cmidrule{2-7}
                                          & \textbf{Hybrid}                  & \multirow{2}{*}{Fine-tuned } & BoW & 51.21 & 90.56 & 65.43 \\
                                          &                                  &                                      & NLI & 83.48 & 71.48 & 77.01
\\
        \bottomrule
    \end{tabular}
    }
    \caption{The results of the lyrics-based, acoustics-based and the hybrid models for the moods ``love'' and ``chill''. }
    \label{tab:case_study}
\end{table*}



\subsection{RQ4. Do Models Capture the Same Information about Lyrics and Acoustics as Humans?}

While the song-mood associations reflected by BNPMI scores are derived from listener playlisting data, there is no way to decompose these scores into associations provided by lyrics, by acoustics, and other factors.

We therefore conduct a human annotation task on 101 songs and 302 (song, mood) pairs, which are selected randomly,  using 3 groups of annotators. The annotation task includes two subtasks: 

\begin{itemize}
    \item \textbf{Lyrics Annotation:} Judge whether a mood is relevant to a song only by reading the lyrics.
    \item \textbf{Acoustics Annotation:} Judge whether a mood is relevant to a song only by listening to the {\em instrumental} version.
\end{itemize}

To obtain `instrumental' versions of the songs with the singing voice -- and hence the lyrics -- removed, we apply vocal source separation on the audio using a U-Net architecture \cite{Jansson2017SingingVS,jansson2019joint}. 

The annotators were given the options of \{Yes, No, Uninformative\} to annotate each (song, mood) pair. The final annotation is decided based on the majority vote, and in a case of disagreement between all three annotators, a fourth annotator resolved the disagreement. Table~\ref{tab:inter_annotator_agreement} shows the inter annotator agreement using Fleiss Kappa~\cite{falotico2015fleiss} ($\kappa$). According to the Fleiss Kappa interpretation table~\cite{landis1977measurement}\footnote{see Appendix ``Fleiss Kappa Interpretation" for the interpretation table}, the annotators have `fair' agreement; a qualitative examination of the annotations suggests that the subtasks are highly subjective. 

Even so, we observe that the degree to which  lyrics and acoustics contribute to the association between songs and mood depends on the specific mood. Consistent with the model predictions, ``chill'' is more inferable from acoustics, while ``love''  is more inferable from lyrics according to the annotations. Moreover, we find  that a subset of the moods in our set are neither inferable from the lyrics nor the acoustics. This subset includes moods like ``minimalist'', ``sunshine'', and ``obsessed''. The association of these moods with songs may be determined by  aspects other than acoustics or lyrics, such as cultural or personal associations. For example, some listeners may playlist a song under the term ``obsessed'' because it is recently popular and those listeners pay attention to the charts; others may do so because that song evokes personal memories and emotions.

 \begin{table}[t]
    \centering
    {
    \begin{tabular}{l c} 
        \toprule
        \textbf{Annotation} & $\kappa$\\
                \midrule

        Lyrics  & 0.2846 \\ 
        Acoustics  & 0.2910 \\
        \bottomrule
    \end{tabular}
    }
    \caption{Inter annotator agreement based on the Fleiss $\kappa$. }
     \label{tab:inter_annotator_agreement}
\end{table}

\begin{table*}[t]
    \centering
  {
    \begin{tabular}{l ccc | ccccc} 
        \toprule
        \textbf{Source of Ground Truth} & Precision & Recall & F1 & TP & TN & FP & FN & uninformative \\
        \midrule
        Lyrics Annotation & 57.28 & 69.09 & 62.63 & 114 & 51 & 85 & 18 & 34 \\
        Acoustics Annotation &  51.23 & 61.53& 55.91 & 104 & 65 & 99
& 15 & 19 \\
        Lyrics \& Acoustics Consensus & 72.01 &	74.40 & 73.19 &	157 & 54 & 61 & 27 & 3 \\
        \bottomrule
    \end{tabular}
    }
    \caption{BNPMI score evaluation against human judgments.}
    \label{tab:bnpmi_score_eval}
\end{table*}

\begin{table*}[t]
    \centering
    \begin{tabular}{lll  ccc | cccc} 
        \toprule
        \textbf{Ground Truth} & \textbf{Feature} & \textbf{Approach}  & Precision & Recall & F1 & TP & TN & FP & FN\\
        \midrule
        \multirow{3}{*}{\textbf{Lyrics Annotation}}    
                                          & \multirow{3}{*}{\textbf{Lyrics}} & Zero-shot                      & 72.58 & 68.18 & 70.31 & 90 & 102 & 34 & 42 \\
                                                                             \cmidrule{3-10}
                                          &                                  & Fine-tuned                      & 58.33 & 90.15 & 70.83 & 119 & 51 & 85 & 13 \\
                                          \cmidrule{2-10}
                                          & \textbf{Hybrid}                  & Fine-tuned                      & 57.00 & 87.02 & 68.88 & 114 & 51 & 85 & 18\\
        \midrule
        \multirow{6}{*}{\textbf{Acoustics Annotation}}   
                                          & \multirow{3}{*}{\textbf{Lyrics}} & Zero-shot                       & 52.62 & 50.42 & 51.50 & 60 & 110 & 54 & 59\\
                                                                             \cmidrule{3-10}
                                          &                                  & Fine-tuned                      & 50.00 & 88.23 & 63.82 & 105 & 59 & 105 & 14\\
                                          \cmidrule{2-10}
                                          & \textbf{Hybrid}                  & Fine-tuned                      & 49.75 & 85.71 & 62.92 & 102 & 60 & 103 & 17\\
        \midrule
        \multirow{6}{*}{\textbf{Lyrics \& Acoustics Consensus}}   
                                          & \multirow{3}{*}{\textbf{Lyrics}} & Zero-shot                     &  79.06 & 55.43 & 65.17 & 102 & 88 & 27 & 82\\
                                                                             \cmidrule{3-10}
                                          &                                  & Fine-tuned                      & 71.87 & 87.50 & 78.92 & 161 & 52 & 63 & 23\\
                                          \cmidrule{2-6}
                                          \cmidrule{2-10}
                                          & \textbf{Hybrid}                  & Fine-tuned                      & 70.90 & 85.24 & 77.41 & 156 & 51 & 64 & 27\\                                
        \bottomrule
    \end{tabular}
    \caption{Evaluation of the NLI-based lyrics and hybird models against the lyrics annotation, acoustics annotation and the consensus of the lyrics and acoustics annotations.}
    \label{tab:model_eval_against_human_annot}
\end{table*}

The listener-generated BNPMI scores are composite  reflections of both lyrics and acoustics (as well as other factors) from the listeners' point of view.
In the previous research questions, we use the BNPMI-derived song-mood association as ground truth to train and evaluate our models. However, in this section, we investigate how much the BNPMI associations are aligned with the human evaluation; i.e., we use the human annotation as ground truth. 

 We define an annotation consensus as follows.
 If at least one source of the annotation, either lyrics or acoustics, annotate the association between the song and mood as positive, the consensus ground truth will be positive association. If one source of the annotation is negative and the other is uninformative, the consensus will be negative. Lastly if both of the sources are uninformative the consensus is uninformative.
 
 Table~\ref{tab:bnpmi_score_eval} shows a comparison of BNPMI-derived song-mood associations against human judgments based on lyrics and acoustics annotations separately and in consensus. The scores demonstrate that lyrics correlate better with BNPMI association scores than acoustics, and that the the BNPMI association scores are most closely consistent with a consensus of lyrics and acoustics. We also see that there are more $(song, mood)$ pairs where the lyrics alone are insufficient for annotators to assign a mood to the track compared to music only. 

\begin{figure}[h]
\centering
\includegraphics[scale=0.44]{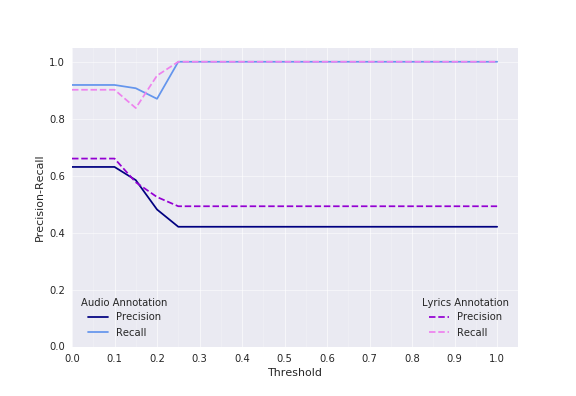}
\caption{Precision Recall vs. Threshold for BNPMI score}
\label{fig:bnpmi_score_thresh_eval}
\end{figure}

Moreover, we investigate the BNPMI threshold set previously to assign the weak association labels by plotting the ``Precision-Recall vs. Threshold" for BNPMI scores. Figure~\ref{fig:bnpmi_score_thresh_eval} shows the precision and recall of BNPMI score evaluation when the true labels are obtained from lyrics annotation as well as audio annotation. By looking at the figure we observe the selected threshold $\tau=0.1$ is the appropriate threshold since the precision and recall are both maximized. 

We further evaluated our models with transformer-based representations (BART model fine-tuned on the MNLI corpus) for lyrics against the lyrics, acoustics and consensus annotations ground truths in Table~\ref{tab:model_eval_against_human_annot}. 
We observe that the zero-shot NLI model captures the association between lyrics and mood very well. Although the difference of the zero-shot NLI and the fine-tuned NLI models in F1 score is negligible, the fine-tuned NLI model  over-predicts  positive associations, resulting in a lower precision. The performance of the Hybrid model is slightly worse than the fine-tuned NLI model, which is the result of the model being more conservative in predicting positive associations.

As expected, we find the zero-shot NLI model does not capture the acoustics, and the fine-tuned NLI model improves the F1 score. The hybrid model  performs slightly worse than the fine-tuned NLI model but the difference in F1 score is negligible.

Lastly, evaluating models against the consensus shows the fine-tuned NLI model captures both the lyrics as well as acoustics. As we expected the performance of the fine-tuned NLI model represented by F1 score improves. 

\section{Discussion}
In this work, we study the association between song lyrics and moods, and compare how much the lyrics and acoustics of a song contribute to understanding the mood of the song. We investigate what lyrics tell us about the song's mood without any training data by formulating the problem as a zero-shot text classification task to classify whether the song's lyrics and the mood are associated with each other. We explore two transformer-based approaches, natural language inference and next sentence prediction for addressing the text-classification task. We find that the natural language inference model is good at capturing literal meaning of the song by having a high precision, however, it has low recall. On the other hand, the next sentence prediction model is biased towards predicting positive associations, resulting in high recall. 

Furthermore, we investigate if training a lyrics-based model on the listener-generated mood associations results in a high performance model to predict the associations for unseen songs. We represent the lyrics with a bag-of-words model as well as transformer-based models. We find that  the fine-tuned transformer-based models outperform the bag of words model, and predict mood associations with high enough precision and  recall that they can be valuable for predicting moods of unseen songs in music applications.

We also compare the relative contributions of lyrics and acoustics to the mood of a song by exploring  models built on each modality, as well as by combining both into a hybrid model. We observe the contribution of lyrics varies depending on the mood. However, both lyrics and acoustics are a source of information for correctly classifying moods, and using a hybrid model trained on both lyrics as well as acoustic results in higher performance. 

Lastly, through a human annotation task, we study whether the models capture the same information about lyrics and acoustics as humans. The annotations demonstrate that understanding the mood of a song by its lyrics or acoustics is a  highly subjective task. However, similar to the lyrics-based and acoustics-based models, we observe that a subset of moods are more inferrable from the lyrics than acoustics and the other way around. Furthermore, we evaluate how much the collaborative listener-generated mood associations are aligned with human judgments, and find that they are mostly aligned with lyrics, suggesting that users seem to pay primary importance to lyrics with naming playlists.

One shortcoming of our combination of lyrics and acoustics in the models is that we use a rich transformer based representation for the former, but summary features with a linear classifier for the latter. As such, the acoustics-based models are best compared on the same footing with the bag-of-words lyrics models rather than the transformer approaches. Future work should include deep models of acoustics for a fair comparison and hybrid combination.

It is worth keeping in mind that the mood expressed in a work of art may differ from the mood experienced by its audience---a picture, a text, or a song which expresses sadness or melancholia may well elicit enjoyment, exhilaration, or admiration in its viewer, reader, or listener \cite{sachs2015pleasures}. An area of future work is to understand whether seemingly incorrect predictions by the models, or contradictions between listener-generated associations and predictions from models using lyrics, arise from these differences of songwriter intent and user perception.

\section{Appendices}
\setcounter{equation}{0}

\subsection{BNPMI Score Calculation}
\label{appendix:bnpmi}

Pointwise mutual information (PMI) is a common measure of association between two possible events. In our context we want to measure the association between songs and words through their occurrence and cooccurrence in playlists of songs. A word is part of a playlist when it is present in the playlist's name or description.  In this way the pointwise mutual information between a
song, s, and a mood, m, is calculated as:

\begin{align}
    \text{PMI}(s,m) = \log\frac{p(s,m)}{p(s)p(m)}
\end{align}

A common normalization of PMI (NPMI) is given by:

\begin{align}
    \text{NPMI}(s, m) = \frac{\text{PMI}(s, m)}{\log p(s, m)}
\end{align}

which has values on [-1, 1], taking on the value -1 if they can never occur
together and 1 one if the can never occur apart. If the events happen independently of one another, it takes on the value 0. It will be convenient to rewrite the formula for NPMI as:

\begin{align}
    \text{NPMI}(s, m) = \frac{\log p(s) - \log p(s|m)}{\log p(m) + \log p(s|m)} \label{eq:npmi}
\end{align}

In order to estimate NPMI, $p(s), p(m), p(s|m)$ is estimated. It is assumed that each mood and song occurs often enough in playlists that empirical estimates of $p(s)$ and $p(m)$ should be ``good enough''.

\begin{align}
    \hat{p}(s) = \frac{\#\text{\{playlists containing }s\}}{\#\text{playlists}} \\
    \hat{p}(m) = \frac{\#\text{\{playlists containing }m\}}{\#\text{playlists}} 
\end{align}

However, there is no assumption that there will be enough cooccurrences of $s$ and $m$ to get good enough estimates of $p(s|m)$. In particular, there might be spurious correlations from “lucky” cooccurences. To account for this there is this assumption that the $p(s|m)$ for a fixed mood $m$ are drawn from a conjugate prior distribution, $\text{Beta}(\alpha_m,\beta_m)$, for some $\alpha_m$ and $\beta_m$. $p_{s,m}$ is denoted to be the empirical estimate of $p(s|m)$ and estimate $\alpha_m$ and $\beta_m$ for each mood $m$ through method of moments as follows:

\begin{align}
    p_{s,m}   &= \frac{ \#\{\text{playlists containing } s \text{ and } m\} } {\#\{\text{playlists containing } m \}} \\
    \Bar{p}_m &:= \frac{1}{\#\text{songs } } \sum_{s} p_{s,m} \\
    \Bar{v}_m &:= \frac{1}{\#\text{songs } - 1} \sum_{s} (p_{s,m} - \Bar{p}_m)^2 \\
    \hat{\alpha} &= \Bar{p}_d \left( \frac{\Bar{p}_m (1 - \Bar{p}_m) }{\Bar{v}_m}  - 1 \right)\\
    \hat{\beta}  &= (1-\Bar{p}_m) \left( \frac{\Bar{p}_m (1 - \Bar{p}_m) }{\Bar{v}_m} - 1 \right)
\end{align}

From this the posterior estimate for $p(s|m)$ is calculated as:
\begin{align}
    \hat{p}(s|m) = \frac{\#\{ \text{playlist containing } s \text{ and } m \} + \hat{\alpha_m}} { \#\{ \text{playlist containing } m \} + \hat{\alpha_m} + \hat{\beta_m }}
\end{align}

The final association score, BNPMI is defined to be the estimate of NPMI that comes from substituting our estimates for $p(s)$, $p(m)$, and $p(s|m)$ into equation~\ref{eq:npmi}:
\begin{align}
    \text{BNPMI}(s,m) = \frac{\log \hat{p}(s) - \log \hat{p}(s|m)}{\log \hat{p}(m) + \log \hat{p}(s|m)} \label{eq:bnpmi}
\end{align}

\begin{table*}
\centering
\begin{tabular}{llcc}
\toprule
\multirow{2}{*}{\textbf{Descriptor}} & \multicolumn{1}{l}{\multirow{2}{*}{\textbf{Song / Artist}}} & \multicolumn{2}{c}{\textbf{Informativeness}}  \\
                            & \multicolumn{1}{c}{} & \textbf{Lyrics} & \textbf{Audio}\\
\midrule
calm  & Stormbringer / deep purple & No  & No \\
sad  & Stormbringer / deep purple  & Yes & No \\
relaxing                    & Stormbringer / deep purple                         & No            & No                   \\
lit                         & Stormbringer / deep purple                         & No            & Yes                  \\
slow                        & Stormbringer / deep purple                         & U & No                   \\
depression                  & Stormbringer / deep purple                         & Yes           & No                   \\
smooth                      & Stormbringer / deep purple                         & No            & No                   \\
chill                       & Stormbringer / deep purple                         & No            & No                   \\
influential                 & Who's Gonna Take The Weight? / Gang Starr          & Yes           & U        \\
sad                         & Who's Gonna Take The Weight? / Gang Starr          & No            & No                   \\
happy                       & Who's Gonna Take The Weight? / Gang Starr          & No            & No                   \\
soft                        & Who's Gonna Take The Weight? / Gang Starr          & No            & No                   \\
militant                    & Who's Gonna Take The Weight? / Gang Starr          & No            & No                   \\
motivation                  & What I'd Say / Earl Thomas Conley                  & No            & No                   \\
upbeat                      & What I'd Say / Earl Thomas Conley                  & No            & No                   \\
chill                       & What I'd Say / Earl Thomas Conley                  & No            & Yes                  \\
good vibes                  & What I'd Say / Earl Thomas Conley                  & No            & Yes                  \\
lit                         & What I'd Say / Earl Thomas Conley                  & No            & No\\
\bottomrule
\end{tabular}
\caption{Examples of mood and song pairs along with  lyrics and audio annotation. U indicates that the judgment source (lyrics or audio) is uninformative.}
\label{tab:mood_song_examples}
\end{table*}

\subsection{Examples of Mood and Song Pairs}
Table~\ref{tab:mood_song_examples} shows 18 pairs of mood and song for 3 different songs along with their lyrics and audio human annotation.

\subsection{Fleiss Kappa Interpretation}
Table~\ref{tab:kappa_interpretation} shows the interpretation of the $\kappa$ value described by~\citet{landis1977measurement}. 

\begin{table}
    \centering
    \scalebox{1}{
    \begin{tabular}{l c} 
        \toprule
        $\kappa$ & interpretation\\
        \midrule
        $<$ 0	        & Poor agreement \\
        (0.01, 0.20]	& Slight agreement \\
        (0.20, 0.40]	& Fair agreement \\
        (0.40, 0.60]	& Moderate agreement \\
        (0.60, 0.80]	& Substantial agreement \\
        (0.80, 1.00]	& Almost perfect agreement \\
        \bottomrule
    \end{tabular}
    }
    \caption{Fleiss $\kappa$ interpretation}
    \label{tab:kappa_interpretation}
\end{table}

\pagebreak
\bibliography{lyricsmood}

\end{document}